\documentclass[pra,a4paper,preprint,showpacs,superscriptaddress]{revtex4}
\usepackage{amsmath}
\usepackage{amsfonts}
\usepackage{graphicx}
\usepackage{longtable}
\newcommand{\be}{\begin{equation}}
\newcommand{\ee}{\end{equation}}

\newcommand{\ba}[1]{\left(\begin{array}{#1}}
\newcommand{\ea}{\end{array}\right)}
\newcommand{\bee}{\begin{equation*}}
\newcommand{\eee}{\end{equation*}}
\begin{document}

\title{Biseparability of noisy  pseudopure, W and GHZ states using conditional quantum relative Tsallis entropy} 
\author{Anantha S Nayak}\affiliation{Department of Physics, Kuvempu University, Shankaraghatta, Shimoga-577 451, India.}
\author{Sudha }
\affiliation{Department of Physics, Kuvempu University, Shankaraghatta, Shimoga-577 451, India.}
\affiliation{Inspire Institute Inc., Alexandria, Virginia, 22303, USA.}
\author{A. R. Usha Devi }
\affiliation{ Department of Physics, Bangalore University, Bangalore 560 056, India}
\affiliation{Inspire Institute Inc., Alexandria, Virginia, 22303, USA.}
\author{A. K. Rajagopal }
\affiliation{Inspire Institute Inc., Alexandria, Virginia, 22303, USA.}
\affiliation{Harish-Chandra Research Institute, Chhatnag Road, Jhunsi, Allahabad-211 019, India.}
\affiliation{Institute of Mathematical Sciences, C.I.T. Campus, Taramani, Chennai-600 113, India.}
\date{\today}
\begin{abstract}
We employ the conditional version of sandwiched Tsallis relative entropy  to determine~$1:N-1$ separability range in the noisy one-parameter families of {\em pseudopure} and {\em Werner-like} $N$-qubit W, GHZ states. The range of the noisy parameter, for which  the conditional sandwiched Tsallis relative entropy is positive, reveals perfect agreement with {\em the necessary and sufficient} criteria for the separability in the $1:N-1$ partition of these one parameter noisy states.         
\end{abstract}
\pacs{03.65.Ud, 03.67.-a} 
\maketitle
\section{Introduction} 
Entropic characterization of separability
in mixed composite states has attracted significant attention~\cite{e1a,e1b,e1c,e1d,e1e,e2a,e2b,tsa,tsb,e3,e4,e5,e6a,e6b,e7a,e7b,e8,e9,asn,asn2}. Nielsen and Kempe~\cite{NK}  brought forth the remarkable feature that  the subsystems of an entangled state may exhibit more disorder than the whole system -- unlike a separable state, which is emphatically more disordered globally than locally. Consequent to this, the von Neumann conditional entropy $S(B\vert A)=S(\rho_{AB})-S(\rho_A)$ of a pure entangled bipartite state is negative. Negative von Neumann conditional entropies would only offer sufficient but  not necessary condition for identifying entanglement in mixed states. For instance,  the two qubit Werner 
state~\cite{wer,pop} expressed as $\rho_{AB}=I_4(1-x)/4+x\,\vert \Psi \rangle \langle \Psi \vert$, $0\leq x \leq 1$, 
$\vert \Psi \rangle=\frac{1}{\sqrt{2}} (\vert 00\rangle+\vert 11\rangle)$, is known to be separable 
in the range $0\leq x \leq \frac{1}{3}$ and entangled in the range 
$1/3 < x \leq 1$.  But positive von-Neumann conditional entropy $S(B\vert A)\geq 0$ results  in the 
separability range $0\leq x \leq 0.747$ for the two-qubit Werner state.   Generalized conditional 
entropies, such as R\'{e}nyi and Tsallis entropies, offer more sophisticated tools to detect entanglement in mixed composite systems~\cite{e1a,e1b,e1c,e1d,e1e,e2a,e2b,tsa,tsb,e3,e4,e5,e6a,e6b,e7a,e7b,e8,e9,asn,asn2}. In fact, the 
conditional version of the Tsallis entropy  
\be
\label{arre}
S^{T}_{q}(A|B)= \frac{1}{q-1} \left[1-  \frac{\mbox{Tr} \rho^{q}_{AB}}{\mbox{Tr} \rho^{q}_{B}}\right], 
\ee
was employed by Abe and Rajagopal~\cite{e3} to obtain the separability range $0\leq x \leq \frac{1}{3}$ 
for the two-qubit Werner state (identified in the limit of $q\rightarrow \infty$ for which  the conditional Tsallis entropy  $S^{T}_{q}(A|B)$ is  positive). The separability criterion using the Abe-Rajagopal q-conditional entropy, ({\emph{AR-criterion}}) was found 
to yield separability ranges matching with the \emph{positivity under partial transpose (PPT) criterion}~\cite{peres,horodecci} in  some one-parameter families of noisy states~\cite{e8}. 

Entropic separability criterion received a further impetus recently with the introduction of the generalized non-commutative conditional sandwiched  Tsallis relative entropy (CSTRE), which is shown~\cite{asn,asn2} to be superior than the Abe-Rajagopal (AR) version of conditional Tsallis entropy in witnessing entanglement.   In fact, the  sandwiched (non-commuting) form of the R\'{e}nyi relative entropy introduced in Refs.~\cite{e11,e12,e13} led to an analogous sandwiched form of the Tsallis relative entropy of a density operator $\rho$ and a positive operator $\sigma$, given by~\cite{asn}, 
\be
\label{cstre1}
\widetilde{D}^{T}_{q}(\rho || \sigma ) = \frac{\mbox{Tr}\left\{
\left(\sigma ^{\frac{1-q}{2q}} \rho \  \sigma ^{\frac{1-q}{2q}}\right)^{q}\right\}-1}{q-1}.
\ee
The Tsallis relative entropy $\widetilde{D}^{T}_{q}(\rho || \sigma )$ is zero if and only if $\rho=\sigma$. 

The new version of the Tsallis relative entropy $\widetilde{D}^{T}_{q}(\rho || \sigma)$  reduces to the traditional relative Tsallis entropy $ D^{T}_{q}(\rho || \sigma ) $
\be
\label{ttre}
D^{T}_{q}(\rho || \sigma ) = \frac{\mbox{Tr}\left[\rho^{q} \ \ \sigma^{1-q}\right]-1} {q-1} 
\ee
when $ \rho $ and $ \sigma $  commute with each other.  

Based on the generalized non-commutative version $\widetilde{D}^{T}_{q}(\rho || \sigma)$ of the Tsallis  relative entropy, one obtains the corresponding  CSTRE to be of the form,  
$ \widetilde{D}^{T}_{q}(\rho_{AB} || I_A\otimes \rho_B )$ of a composite bipartite state $\rho_{AB}$  and the positive operator $I_A\otimes \rho_B$    (where $\rho_B$ is the subsystem state $\rho_B={\rm Tr}_A[\rho_{AB}]$, and  $I_A$ denotes the identity matrix in the Hilbert space of the subsystem A)  as~\cite{asn}, 
\be
\label{tcsre}
\widetilde{D}^{T}_{q} \left( \rho_{AB} || I_A\otimes \rho_{B} \right) = \frac{\widetilde{Q}_{q} \left( 
\rho_{AB} || I_A\otimes \rho_{B} \right) - 1 } {1-q}.
\ee
Here, we have denoted,  
\[
\widetilde{Q}_{q} \left( \rho_{AB} || I_A\otimes \rho_{B} \right) 
= \mbox{Tr} \left\{\left[\left(I_{A} \otimes 
\rho_{B}\right)^{\frac{1-q}{2q}} \rho_{AB} \ \left(I_{A} \otimes \rho_{B}\right)^{\frac{1-q}{2q}} 
\right] ^{q} \right\} = \sum_{i} \,\lambda_i^q,
\]  
where $\lambda_i$ are the eigenvalues of 
the sandwiched matrix $\left(I_{A} \otimes 
\rho_{B}\right)^{\frac{1-q}{2q}} \rho_{AB} \ \left(I_{A} \otimes \rho_{B}\right)^{\frac{1-q}{2q}}$. Non positive values of the CSTRE $\widetilde{D}^{T}_{q} \left( \rho_{AB} ||I_A\otimes \rho_{B} \right)$ with $q>1$, i.e., 
\be 
\label{imp}
\widetilde{D}^{T}_{q} \left( \rho_{AB} ||I_A\otimes \rho_B \right)=\frac{\left(\sum_{i} \,\lambda_i^q\right)-1}{1-q}<0
\ee 
imply entanglement (see Ref.~\cite{asn,asn2}).  

When the subsystem $ \rho_{B} $ is maximally mixed,  the CSTRE $\widetilde{D}^{T}_{q} \left( \rho_{AB} ||I_A\otimes \rho_{B} \right)$  reduces to the commutative version viz., the 
AR q-conditional Tsallis entropy $S^{T}_{q}(A|B)$ of (\ref{arre}).  
In our earlier papers~\cite{asn,asn2} we had investigated bipartite separability of one parameter noisy 
symmetric multiqubit systems based on the non-positivity of both AR  conditional entropy  and the 
corresponding CSTRE; and we had shown that whenever the subsystem is not maximally mixed,  the CSTRE 
criterion yields stricter separability range than that obtained through the commutative AR version. 
In this article, we extend the  CSTRE criterion to witness entanglement in noisy one parameter families 
of the $N$-qubit pseudopure states~\cite{ec} and the $N$-qubit generalizations of Werner-like 
one parameter states~\cite{wer} involving  W, GHZ states. 
We show that the non-commutative CSTRE criterion is both necessary and sufficient to detect 
entanglement in the  $(1:N-1)$ partitions of the one parameter noisy multiqubit states 
explored here.  
 %%%%%%%%%%%%%%%%%%%%%%%%%%%%%%%%%%%%%   
\section{Pseudopure $N$ qubit W and  GHZ states} 
%%%%%%%%%%%%%%%%%%%%%%%%%%%%%%%%%%%%%%%%%%%
The  pseudopure (PP) families of states are formed by mixing any pure state with  white noise~\cite{ec}. They have played a crucial role in  demonstrating quantum information processing possibilities in liquid state NMR spectroscopy~\cite{nmr1,nmr2}. In Ref.~\cite{ec}, different measures of quantum correlations  of bipartite $d\times d$ PP states of the form    
\begin{eqnarray}
\label{pp}
\rho^{\rm PP}_{\phi}(x)=\frac{1-x}{d^2 -1 } \left[\left( I_d \otimes I_d \right)- \vert \phi\rangle \langle \phi\vert \right] + x \vert \phi \rangle \langle \phi\vert
\end{eqnarray}
(where $\vert \phi\rangle$ is any arbitrary $d\times d$ pure entangled state and  $0\leq x\leq 1$ denotes the noisy parameter) are examined. Here we  investigate entanglement in the  $(1:N-1)$ bipartition of the $N$ qubit PP states, constructed using W and GHZ states, based on the CSTRE approach.  

The one parameter family of $N$-qubit pseudopure states
\begin{eqnarray}
\label{ppw}
\rho^{\rm PP}_{{\rm W}_N}(x)=\frac{1-x}{2^N -1 }\, \left( I_2^{\otimes N}- \vert {\rm W}_N\rangle \langle {\rm W}_N\vert \right) + x \vert {\rm W}_N \rangle \langle {\rm W}_N\vert \nonumber 
\end{eqnarray}
 obtained by considering the pure state $\vert \phi \rangle$ in (\ref{pp})  to be the 
$N$-qubit W state:  
\begin{eqnarray}
\label{ws}
\left| {\rm W}_N \right\rangle  & = &  \frac{1}{\sqrt{N}} [\left|1_1 0_2 \cdots 0_N\right\rangle 
+ \left|0_1 1_2\cdots 0_{N}\right\rangle +  \cdots+  \cdots + \left|0_1 0_2 0_3\cdots 1_N\right\rangle] 
															\end{eqnarray}
and the $d\times d$ matrix $I_d \otimes I_d$ replaced by its multiqubit counterpart $I_2^{\otimes N}$.   

We focus on finding the $1:N-1$ separability range of the W family of PP states $\rho^{\rm PP}_{{\rm W}_N}(x)$  using CSTRE 
criterion. For this purpose, an evaluation of  
the eigenvalues $\lambda_i(x)$  of the sandwiched matrix  
\[
\left(I_2 \otimes 
\sigma^{\rm PP}_{{\rm W}_{N-1}}(x)\right)^{\frac{1-q}{2q}} \rho^{\rm PP}_{{\rm W}_N}(x) \ \left(I_2 \otimes 
\sigma^{\rm PP}_{{\rm W}_{N-1}}(x)\right)^{\frac{1-q}{2q}},
\] 
where $\sigma^{\rm PP}_{{\rm W}_{N-1}}(x)={\rm Tr}_1[\rho^{\rm PP}_{{\rm W}_N}(x)]$ denotes the $N-1$ qubit subsystem of $\rho^{\rm PP}_{{\rm W}_N}(x)$, needs to be carried out. 
We obtain the following explicit structure of the eigenvalues $\lambda_i$   (for $N\geq 3$):    
\begin{eqnarray}
\label{epsilon}
\nonumber \lambda_{1}&=&\left(2\right)^\frac{1-q}{q} \left(\frac{1-x}{2^{N}-1}\right)^\frac{1}{q}, \ \  \mbox{ $\left(2^{N} - 4\right)$ fold-degenerate}; \\
\nonumber \lambda_{2} & = & \left(\frac{1-x}{2^{N} - 1}\right) \left[\frac{ (2N -1) +\left(\sum^{N}_{j=3} 2^{j-1} - 2(N-2)\right)x}{ N  \left( 2^{N} - 1\right)}\right]^\frac{1-q}{q},  
\\
 \lambda_{3} &=& \left(\frac{1-x}{2^{N} - 1}\right)  \left[\frac{ (N + 1) +\left(\sum^{N}_{j=3} 2^{j - 1} + (N-2) \left( 2^{N} - 2\right)\right)x}{ N \left( 2^{N} - 1\right)}\right]^\frac{1-q}{q}, \\
\nonumber \lambda_{4/5}&=& \left[N \left( 2^{N} - 1\right)\right]^{\frac{-1}{q}}  \left( \frac{1}{2}\right) \left[ \alpha \ a + \beta \ b \pm \sqrt{ \left(\alpha \ a + \beta \ b \right)^{2} + 8 N^{2} (2^{N}- 1) x (x - 1)\  \alpha \ \beta } \right].
\end{eqnarray}
 where 
\begin{eqnarray}
\nonumber  \alpha & = & \left[ 2N - 1 + \left(\sum^{N}_{j=3} 2^{j-1} - 2 (N-2)\right)x\right]^\frac{1-q}{q}, \\ 
\nonumber \beta &=& \left[ N + 1 + \left(\sum^{N}_{j=3} 2^{j-1} + (N - 2) \left( 2^{N} - 2\right) \right) x \right]^\frac{1-q}{q}, \\
 a & = &  \left(N-1\right) +\left(\sum^{N}_{j=3} 2^{j-1} - N + 4 \right) x,  \\ 
\nonumber b &=& 1 + \left( \sum^{N}_{j=3} 2^{j-1} + (N - 2) \left(2^{N} - 2\right) + N \right) x.
\end{eqnarray}
\begin{table}[h] 
\begin{center}
\caption{The comparison of the $1:N-1$ separability range of the state 
$\rho^{\rm PP}_{{\rm W}_N} (x)$, for $N = 3, \ 4,\ 5,\ 6 $ obtained through different separability criteria.} 
\label{ppw1}
\begin{tabular}{|c|c|c|c|c|}
\hline
& & & & \\
Number & von Neumann & AR  &  CSTRE & PPT     \\ 
 of & conditional  & q-conditional &  &   \\
qubits ($N$) & entropy & entropy &  &   \\
\hline\hline 
3  & 0.7390 & 0.3636 & 0.3083 & 0.3083 \\ \hline
4  & 0.6963 & 0.25   & 0.1807 & 0.1807 \\ \hline 
5  & 0.6723 & 0.1621 & 0.1014 & 0.1014 \\ \hline 
6  & 0.6621 & 0.1    & 0.0552 & 0.0552 \\ \hline 
\end{tabular}
\end{center}
\end{table}

Substituting these eigenvalues $\lambda_i$ in (\ref{imp}), a numerical estimation of the $1:N-1$ CSTRE separability range for $N=3,\,4,\,5,\,6$ has been carried out. This results in the separability range for the noisy parameter $x$  to be $(0, \ 0.3083)$, $(0, \ 0.1807)$, $ (0,\ 0.1014)$, $ (0,\ 0.0552)$ in the $1:2$, $1:3$, $1:4$, $1:5$ partitions of the noisy state $\rho^{\rm PP}_{{\rm W}_N}(x)$ with $N=3$, $N=4$, $N=5$, $N=6$ respectively.
The results obtained based on the CSTRE, along with the corresponding cut-off value of the parameter $x$  obtained using the AR- and the PPT criteria are listed in Table~\ref{ppw1}.  This offers a direct comparison of different approaches, each leading to the threshold values of the parameter $x$ (beyond which the noisy state is found to be entangled).   
From  Table~\ref{ppw1}  it is clearly seen that, for the noisy state $\rho^{\rm PP}_{{\rm W}_N}(x)$, CSTRE  
provides better separability range than the AR-criterion. Moreover, the CSTRE separability range 
matches identically with the PPT separability range. 

\normalsize In general, the CSTRE criterion (the inequality (\ref{imp}) in the limit $q\rightarrow \infty$) leads to,   
\be
\label{nwsr}
0\leq x \leq \frac{N + \sqrt{N - 1}}{ N +2^{N} \, \sqrt{N-1}}
\ee
for the separability range in the  $(1:N-1)$ partition of the noisy $N$ qubit PP state $\rho^{\rm PP}_{{\rm W}_N}(x)$ for $N\geq 3$. 
Alternately, in the parameter region 
\[
\frac{N + \sqrt{N - 1}}{ N +2^{N} \, \sqrt{N-1}}<x\leq 1,
\] 
the CSTRE method witnesses entanglement in the $(1:N-1)$ bipartition of the noisy state. 

The PP family of states (see (\ref{pp})) with the pure entangled state $\vert\phi\rangle$ expressed in terms of the Schmidt co-efficients i.e.,  $\vert\,\phi\,\rangle=\sum_{i=1}^{d}\, u_i\, \vert \, i_A i_B\rangle$, with $u_1\geq u_2\geq \cdots \geq u_d\geq 0$ are shown to be separable iff~\cite{vidaltarnach,ec}    
\be
\label{vt1}
0\leq x \leq \frac{1+u_1 u_2}{1+d^2\, (u_1 u_2)} 
\ee
For the PP state $\rho^{\rm PP}_{{\rm W}_N}(x)$ of (\ref{ppw}) with  $(1:N-1)$ bipartition under investigation, the Schmidt coefficients (positive square roots of the eigenvalues of the reduced single qubit subsystem density matrix) of the $N$ qubit $W$ state are given by, 
\be
\label{vtw}
u_1=\sqrt{\frac{N-1}{N}}, \ \  u_2=\frac{1}{\sqrt{N}}.   
\ee
Substituting (\ref{vtw}) and replacing $d^2$ by $2^N$ in (\ref{vt1}), we recover the result  (\ref{nwsr}) for the separability range. 
This establishes that the CSTRE approach serves as both necessary and sufficient to detect entanglement in the $(1:N-1)$ partition of the PP state $\rho^{\rm PP}_{{\rm W}_N}(x)$.    

We now proceed to investigate the noisy one parameter family of $N$ qubit PP states $\rho^{\rm PP}_{{\rm GHZ}_N} (x)$ given by, 
\begin{eqnarray}
\label{ppghz}
\rho^{\rm PP}_{{\rm GHZ}_N}(x)=\frac{1-x}{2^N -1 }\, \left( I_2^{\otimes N}- \vert {\rm GHZ}_N\rangle \langle {\rm GHZ}_N\vert \right) + 
x \vert {\rm GHZ}_N \rangle \langle {\rm GHZ}_N\vert. \nonumber 
\end{eqnarray}
where, 
\be
\label{gs}
\vert {\rm GHZ}_N \rangle=\frac{1}{\sqrt{2}} \left( \vert 0_10_2\cdots 0_N \rangle+ 
\vert 1_1 1_2\cdots 1_N \rangle   \right)  
\ee

To find  the $1: N-1$ separability range of $\rho^{\rm PP}_{{\rm GHZ}_N}(x)$ using CSTRE approach, one needs to  evaluate the eigenvalues $\lambda_i$ of the sandwiched matrix 
$\left(I_2 \otimes \sigma^{PP}_{{\rm GHZ}_{N-1}}(x)\right)^{\frac{1-q}{2q}} \rho^{\rm PP}_{{\rm GHZ}_N}(x) \ \left(I_2 \otimes 
\sigma^{\rm PP}_{{\rm GHZ}_{N-1}}(x)\right)^{\frac{1-q}{2q}}$, where $\sigma^{\rm PP}_{{\rm GHZ}_{N-1}}(x)={\rm Tr}_1[\rho^{\rm PP}_{{\rm GHZ}_N}(x)]$ corresponds to the $N-1$ qubit subsystem of $\rho^{\rm PP}_{{\rm GHZ}_N}(x)$. The  non-zero  eigenvalues $\lambda_i$ are given below (for $ N \geq 3 $) in  
(\ref{epsilon2}):
\begin{eqnarray} 
\label{epsilon2}
\nonumber \lambda_{1} & = &\left[\frac{1-x}{2^{N} - 1}\right] \left[
\frac{ 2\left(1-x\right)}{2^{N} - 1}\right]^\frac{1-q}{q}; \ \mbox{($2^{N}-4)$-fold degenerate} \\
\lambda_{2} & = & \left[\frac{1-x}{2^{N} - 1}\right]\left[\frac{ 3 +\left( \sum^{N}_{j = 3} 2^{j - 1}\right)x}{\sum^{N}_{j = 1} 2^{j}}\right]^\frac{1-q}{q};\ \mbox{$3$-fold degenerate}
 \nonumber \\
\lambda_{3} & = & x \left[\frac{ 3 +\left( \sum^{N}_{j = 3} 2^{j - 1}\right)x}{  \sum^{N}_{j = 1} 2^{j}}\right]^\frac{1-q}{q}.
\end{eqnarray}

Substituting these eigenvalues $\lambda_i$ in (\ref{imp}), we numerically evaluate the $1:N-1$ separability range (beyond which the CSTRE is negative and hence imply entanglement) for specific cases $N=3,\,4,\,5,\,6$. We obtain the result  $ [0,\, 0.3], [0,\, 0.1666], [0,\, 0.0882], [0,\, 0.0454] $  
as the separability ranges for the noisy state $\rho^{\rm PP}_{{\rm GHZ}_N}(x)$ in its $1:2$, $1:3$, $1:4$, $1:5$ partitions with $N=3,\,4,\,5,\,6$ respectively. We verify that  these results agree with the ones obtained based on both AR and PPT criteria. It may however be identified that though the CSTRE and AR criteria result in the same separability threshold for the noisy parameter $x$,  they approach the cut-off value with different convergence rates, which is depicted in Fig. 1, for the specific case of $N=6$.  
\begin{figure}[ht]
\label{fig1}
\begin{center}
\includegraphics*[width=4in,keepaspectratio]
{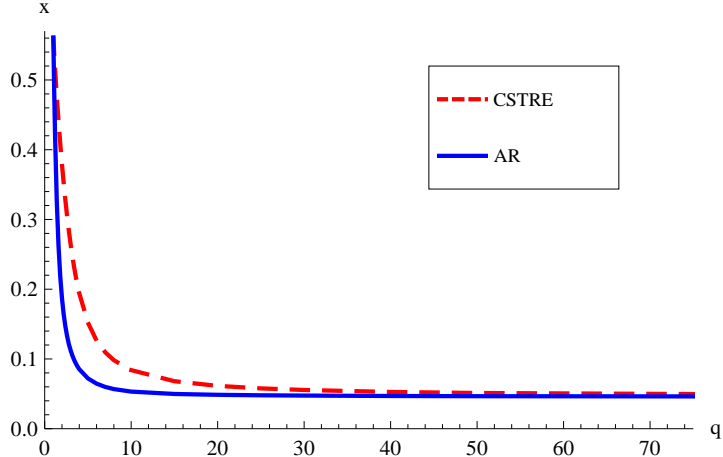} 
\caption{(Color Online) Implicit plots of 
$\tilde{D}^{T}_q(\rho^{\rm PP}_{{\rm GHZ}_N}\vert\vert I_2\otimes\sigma^{\rm PP}_{{\rm GHZ}_{N-1}})=0$  (dashed line) and the Abe-Rajagopal $q$-conditional entropy $S_q^{T}(A\vert B)=0$ 
(solid line) as a function of 
$q$ in the $1:5$ partition of the state $\rho_{{\rm GHZ}_6}^{\rm PP}(x)$. This demonstrates 
the relatively slower convergence of the noisy parameter $x$ to the cut-off value $0.04545$ in the case of the CSTRE approach, when compared with that of the AR method.  
(The quantities plotted are dimensionless).}  
\end{center}
\end{figure}

In general for any $N\geq 3$, we obtain the  following  bound  
 \be 
\label{nggsr}
0\leq x \leq \frac{3}{ 2^{N} + 2 }
\ee
in the limit $q\rightarrow \infty$, within which the  PP state 
$\rho^{\rm PP}_{{\rm GHZ}_N}(x)$ is separable.  

This result matches identically  with the necessary and sufficient condition (\ref{vt1}) for separability (obtained by substituting the Schmidt coefficients associated with the $(1:N-1)$ partition of the GHZ state i.e., $u_1=u_2=1/\sqrt{2}$). Thus, the CSTRE method  is found to serve as a necessary and sufficient condition to detect entanglement in the $1:N-1$ partition of the $N$ qubit PP state $\rho^{\rm PP}_{{\rm GHZ}_N}(x)$.   
%%%%%%%%%%%%%%%%%%%%%%%%%%%%%%%%%%%%%%%%%%%%%%%%
\section{Werner-like one parameter noisy families of $N$ qubit W and GHZ states} 
%%%%%%%%%%%%%%%%%%%%%%%%%%%%%%%%%%%%%%%%%%%%%%%%%%
We consider  
the $N$-qubit generalizations of  Werner-like one parameter noisy family of states 
\begin{eqnarray}
\label{nsnoisy}
\rho_{\Phi_N} (x) &=&(1-x) \frac{I_2^{\otimes N}}{2^N} + \,x \, \left\vert \Phi \right\rangle\left\langle \Phi\right\vert, \ \ \ \ 0\leq x \leq 1 
\end{eqnarray} 
and investigate the separability in the $(1:N-1)$ partition based on CSTRE approach~\cite{footnote}.   
When the pure entangled state $\vert \Phi\rangle$  corresponds to the $N$-qubit W state (See  (\ref{ws})), we get the noisy state 
\begin{eqnarray} 
\label{noisyw}
 \rho_{{\rm W}_{N}} (x) & = & (1-x) \frac{I_2^{\otimes N}}{2^N} + \  x \ \left|{\rm W}_N \right\rangle\left\langle {\rm W}_N\right|. \ \ \ 
\end{eqnarray} 
In order to carry on the task of identifying the $1:N-1$ separability range of the state $\rho_{{\rm W}_{N}}(x)$ via the CSTRE method, we evaluate the $2^N$  eigenvalues $\lambda_i$ of the `sandwiched' matrix $(I_2\otimes \sigma_{{\rm W}_{N-1}})^{\frac{1-q}{2q}}\rho_{{\rm W}_{N}}(x)(I_2 \otimes \sigma_{{\rm W}_{N-1}})^{\frac{1-q}{2q}}$ with 
$\sigma_{{\rm W}_{N-1}}(x)={\rm Tr}_1[\rho_{{\rm W}_N}(x)]$ and they are given by 
\begin{eqnarray}
\nonumber \lambda_{1} & = &
\left(\frac{1-x}{2^{N}}\right) 
\left[\frac{1-x}{2^{N-1}}\right]^\frac{1-q}{q}; 
\ \ \ \ \ \ \ \ \  \mbox{ ($2^{N} - 4)$ fold-degenerate} \\
\nonumber \lambda_{2} & = & 
\left(\frac{1-x}{2^{N}}\right) 
\left[\frac{ N +\left(\sum^{N}_{j=3} 2^{j-2} - 
(N-2)\right)x}{ N \ 2^{N-1}}\right]^\frac{1-q}{q}; 
\\
 \lambda_{3} & = & 
\left(\frac{1-x}{2^{N}}\right) 
\left[\frac{ N +\left(\sum^{N}_{j=3} 2^{j-2} + (N-2) \left( 2^{N-1} - 1\right)\right)x}{ N \ 2^{N-1}}\right]^\frac{1-q}{q};\\
\nonumber \lambda_{4/5} & = & 
\frac{1}{4}\left( 2^{N-1} \,N\right)^{\frac{-1}{q}}  
 \left[ \alpha \ a + 
\beta \ b \pm \sqrt{ \left(\alpha \ a - \beta \ b 
\right)^{2} + 2^{2N + 2} (N - 1) x^{2} \  \alpha \ 
\beta}\right] 
\end{eqnarray}
where 
\begin{eqnarray}
  \alpha & = & \left[ N +\left(\sum^{N}_{j=3} 2^{j-2} - (N-2)\right)x\right]^\frac{1-q}{q},\nonumber\\
\beta &=& \left( N + \left[\sum^{N}_{j=3} 2^{j-2} + (N-2) (2^{N-1}- 1 )\right) x \right]^\frac{1-q}{q}, \nonumber\\
a & = &  N +\left(\sum^{N}_{j=3} 2^{j-2} - (N-2) + 2^{N-1}\right) x, \\ 
b &=& N + \left( \sum^{N}_{j=3} 2^{j-2} + 2^{N-1}(2N-3) - (N-2)\right) x.\nonumber
\end{eqnarray}

Substituting these eigenvalues in (\ref{imp}), we numerically estimate the separability ranges 
in the $1:2$, $1:3$, $1:4$, $1:5$ bipartitions of the noisy states $\rho_{{\rm W}_{3}}(x)$, $\rho_{{\rm W}_{4}}(x)$, $\rho_{{\rm W}_{5}}(x)$, $\rho_{{\rm W}_{6}}(x)$ respectively. 
 We have tabulated (see  Table~\ref{sepr}) the separability threshold value of the parameter $x$ obtained using CSTRE approach, along with the corresponding results from  PPT criteria and also those inferred via the positivity of the corresponding  von Neumann and the AR-conditional entropies.      
\begin{table}[h] 
\begin{center} 
\caption{ The $1:N-1$ separability threshold value of the noisy parameter $x$ in  the states $\rho_{{\rm W}_{N}} (x)$ for $N = 3,\ 4,\ 5,\ 6 $,  obtained via the positivity of the CSTRE, the von Neumann and the AR conditional entropies, along with the one obtained from the PPT criteria.}
\label{sepr}

\begin{tabular}{|c|c|c|c|c|}
\hline
& & & & \\
Number & von Neumann & AR  &  CSTRE & PPT     \\ 
 of & conditional  & q-conditional &  &   \\
qubits  & entropy & entropy &  &   \\
\hline\hline 
3  & 0.7018 & 0.2727 & 0.2095 & 0.2095 \\ \hline
4  & 0.6760 & 0.2    & 0.1261 & 0.1261 \\ \hline 
5  & 0.6618 & 0.1351 & 0.0724 & 0.0724 \\ \hline 
6  & 0.6567 & 0.0857 & 0.0402 & 0.0402 \\ \hline 
\end{tabular}
\end{center}
\end{table}
It is readily seen that the result based on the positivity of the CSTRE is stronger than the one obtained from the positivity of the von Neumann, AR conditional entropies. Further, it is observed that the CSTRE result agrees with that identified from the PPT criterion.    

In general, the CSTRE approach is found to lead to the  separability range 
\be
\label{wwsr}
0\leq x \leq \frac{N}{N + 2^{N} \ \sqrt{N - 1}}
\ee
for the $1:N-1$ partitions of the state $\rho_{{\rm W}_{N}}(x)$ for  $N\geq 3$.  
We recall that the noisy $N$-qubit state $\rho_{\Phi_N}(x)$ of (\ref{nsnoisy}) is known to be separable iff~\cite{vidaltarnach} 
\begin{equation} 
\label{nx}
0\leq x\leq \frac{1}{2^N\, u_1\, u_2+1}
\end{equation}  
where $u_1$ and $u_2$ are the two largest Schmidt coefficients of the pure entangled state $\vert\Phi_N\rangle$ under bipartition.
In the specific case of $(1:N-1)$ partition of the state $\rho_{{\rm W}_{N}}(x)$, on substituting  the corresponding Schmidt coefficients
 (see (\ref{vtw})) $u_1=\sqrt{\frac{N-1}{N}}, \ \  u_2=\frac{1}{\sqrt{N}}$  in  (\ref{nx}), one can recognize that  \
the separability range reveals a clear agreement with (\ref{wwsr}) obtained via the CSTRE approach. This establishes that the CSTRE method serves as  necessary and sufficient for inferring separability in this example too. 

We continue to investigate the separability in the $(1:N-1)$ partition of the noisy Werner-like $N$ qubit GHZ state
\begin{eqnarray} 
\label{noisyghz}
 \rho_{{\rm GHZ}_{N}} (x) & = & (1-x) \frac{I_2^{\otimes N}}{2^N} + \  x \ \left|{\rm GHZ}_N \right\rangle\left\langle {\rm GHZ}_N\right| \ \ \ 
\end{eqnarray} 
using CSTRE criteria. Here, the eigenvalues of the sandwiched matrix $(I_2\otimes \sigma_{{\rm GHZ}_{N-1}})^{\frac{1-q}{2q}}\rho_{{\rm GHZ}_{N}}(x)(I_2 \otimes \sigma_{{\rm GHZ}_{N-1}})^{\frac{1-q}{2q}}$, with $\sigma_{{\rm GHZ}_{N-1}}(x)={\rm Tr}_1[\rho_{{\rm GHZ}_N}(x)]$, for any $N\geq 3$ are found to be 
\begin{eqnarray}
\label{nghzeig}
\nonumber \lambda_{1} & = &\left[\frac{1-x}{2^{N}}\right] \left[\frac{1-x}{2^{N-1}}\right]^\frac{1-q}{q}; \ \mbox{  $ (2^{N}-4) $-fold degenerate}; \\
\nonumber \lambda_{2} & = & \left[\frac{1-x}{2^{N}}\right] \left[\frac{ 1 +\left( 2^{N-2} - 1 \right)x}{  2^{N-1}}\right]^\frac{1-q}{q}; \ \mbox{ $3$-fold degenerate} \nonumber \\
 \lambda_{3} & = & \left[\frac{ 1 +\left( 2^{N} - 1 \right)x}{  2^{N}}\right] \left[\frac{ 1 +\left( 2^{N-2} - 1 \right)x}{  2^{N-1}}\right]^\frac{1-q}{q}. 
\end{eqnarray}
Substituting (\ref{nghzeig}) in (\ref{imp}) we find that positivity of $CSTRE$ as $q\rightarrow\infty$ requires the following bounds    
\be
\label{gwsr}
0\leq x \leq \frac{1}{2^{N - 1} + 1}.  
\ee
on the noisy parameter $x$. This result agrees with the  $1:N-1$ separability range obtained based on the commutative AR method too in the case of $\rho_{{\rm GHZ}_{N}}(x)$.  However, {\emph the convergence towards the threshold value of the parameter $x\rightarrow \frac{1}{ 2^{N - 1} + 1 }$ in the  limit $q\rightarrow \infty$  based on the CSTRE method is slower compared to that of the  AR approach. This is illustrated in Fig. 2 in the specific case of $N=6$.  
\begin{figure}[ht]
\begin{center}
\includegraphics*[width=4in, keepaspectratio]
{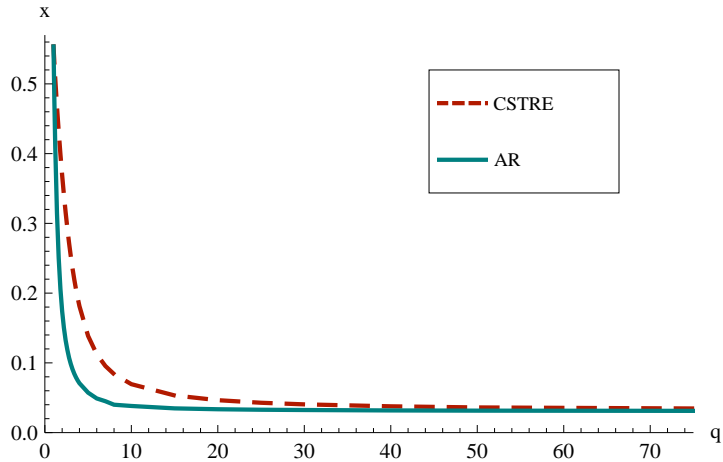} 
\label{wernerghz}
\caption{(Color Online) Implicit plots of 
$\tilde{D}^{T}_q(\rho_{{\rm GHZ}_6}\vert\vert 
I_2\otimes\sigma_{{\rm GHZ}_5})=0$ as a function of 
$q$ (dashed line) and the AR 
$q$-conditional entropy $S_q^{T}(A\vert B)=0$ 
(solid line) for 
$\rho_{{\rm GHZ}_6}(x)$ in its $1:5$ partition. 
The convergence of the parameter 
$x$ to its bound $0.0303$ under the CSTRE criterion is slower compared to that of the AR method.  
(The quantities plotted are dimensionless).}  
\end{center}
\end{figure}
Moreover, substituting the Schmidt coefficients $u_1=u_2=1/\sqrt{2}$ associated with the $(1:N-1)$ partition of the GHZ state in  (\ref{nx}), reveals that  the  range ({\ref{gwsr}}) for the parameter $x$
  obtained from CSTRE approach is both necessary and sufficient for the separability in the $(1:N-1)$ bipartition of the state 
	$\rho_{{\rm GHZ}_{N}}(x)$.

\section{Conclusion} 
We have evaluated the $1 : N - 1$ separability range in the noisy $N$ qubit states of the  
 PP, W and GHZ family using the CSTRE approach.  
Our results show that the positivity of the CSTRE in the limit $q\rightarrow \infty$ is both 
necessary and sufficient criterion for the separability of the $(1:N-1)$ partition of the  
one parameter family of noisy PP, W and GHZ states.  

\section*{Acknowledgment:}  
Anantha S. Nayak acknowledges the support of Department of Science and Technology (DST), Govt. of India through the 
award of INSPIRE fellowship; A. R. Usha Devi is supported under the University Grants Commission (UGC), India (Grant No. MRP-MAJOR-PHYS-2013-29318).  
 

\begin{thebibliography}{88}
\bibitem{footnote} Note that both PP and Werner-like noisy one parameter family of states considered here are not permutation symmetric states, as they do not get restricted only to the $N+1$ dimensional symmetric subspace of the $2^N$ Hilbert space of  $N$-qubits. This is because the completely random  state $I_2^{\otimes N}/2^N$ (which is mixed with the pure symmetric $N$ qubit states in  $\rho^{\rm PP}_{{\rm W}_N}(x)$, $\rho^{\rm PP}_{{\rm GHZ}_N}(x)$, $\rho_{{\rm W}_{N}} (x)$, $\rho_{{\rm GHZ}_{N}}(x)$) is not confined to the $N+1$ dimensional subspace of permutation symmetric $N$ qubit states. In Refs. \cite{asn,asn2} noisy one parameter families of symmetric $N$ qubit states were investigated using  AR- and CSTRE criteria.
\end{thebibliography}
\end{document}